\documentclass[aps,twocolumn,superscriptaddress,altaffilletter,lengthcheck, tightenlines,showpacs,showkeys]{revtex4}

\newcommand{\al}{\alpha}
\newcommand{\bt}{\beta}
\newcommand{\D}{\Delta}
\newcommand{\de}{\delta}
\newcommand{\ben}{\begin{eqnarray}}
\newcommand{\een}{\end{eqnarray}}
\newcommand{\be}{\begin{equation}}
\newcommand{\ee}{\end{equation}}
\newcommand{\ba}{\begin{eqnarray}}
\newcommand{\ea}{\end{eqnarray}}
\newcommand{\n}{\label}

\newcommand{\ep}{\epsilon}
\newcommand{\ga}{\gamma}
\newcommand{\ro}{\rho}

\newcommand{\bn}{\begin{equation}\label}

\newcommand{\tbb}{t_{_{BB}}}

\newcommand{\qmm}{Q^{^{(+,+)}}}
\newcommand{\qme}{Q^{^{(+,-)}}}
\newcommand{\qem}{Q^{^{(-,+)}}}
\newcommand{\qee}{Q^{^{(-,-)}}}

\newcommand{\aee}{a^{^{(-,-)}}}
\newcommand{\aem}{a^{^{(-,+)}}}
\newcommand{\amm}{a^{^{(+,+)}}}
\newcommand{\ame}{a^{^{(+,-)}}}

\usepackage[dvipdf]{epsfig}
\usepackage{color}

\begin{document}

\title{Finite time future singularities in the interacting dark sector  }

\author{Mauricio Cataldo}\email{mcataldo@ubiobio.cl}
\affiliation{Departamento de F\'{\i}sica, Facultad de Ciencias, Universidad del
B\'{\i}o-B\'{\i}o, Avenida Collao 1202, Casilla 5-C, Concepci\'on, Chile}
\author{Luis P. Chimento}\email{chimento@df.uba.ar}
\affiliation{Departamento de F\'{\i}sica, Facultad de Ciencias,
Universidad del B\'{\i}o-B\'{\i}o, Avenida Collao 1202, Casilla 5-C,
Concepci\'on, Chile} \affiliation{Departamento de F\'{\i}sica,
Facultad de Ciencias Exactas y Naturales, Universidad de Buenos
Aires and IFIBA, CONICET, Ciudad Universitaria, Pabell\'on I, Buenos
Aires 1428, Argentina}
\author{Mart\'{\i}n G. Richarte}\email{martin@df.uba.ar}
\affiliation{Departamento de F\'isica, Universidade Federal do Paran\'a,\\ 
Caixa Postal 19044, 81531-990 Curitiba, Brazil}
\affiliation{Departamento de F\'{\i}sica, Facultad de Ciencias Exactas y Naturales, Universidad de Buenos Aires and IFIBA, CONICET, Ciudad Universitaria, Pabell\'on I, Buenos
Aires 1428, Argentina}

\bibliographystyle{plain}

\begin{abstract}

We construct a piecewise model that gives a physical viable
realization of finite-time future singularity  for a spatially flat
Friedmann-Robertson-Walker universe within the interacting dark
matter--dark energy framework, with the latter one  in the form of
a variable vacuum energy. The scale factor solutions provided by the model are
accommodated in several branches defined in four regions delimited
by the scale factor and the effective energy density. A branch
starts from a big bang singularity and describes an expanding
matter-dominated universe until the sudden future singularity
occurs. Then, an expanding branch emerges from a past singularity,
reaches a maximum, reverses its expansion and possibly collapses into itself
while  another expanding branch emerges from the latter singularity and
has a stable de Sitter phase which is intrinsically stable. We
obtain a different piecewise scale factor which describes a contracting de
Sitter universe in the distant past until the finite-time future
singularity happens. It emerges and continues in a contracting
phase, bounces at the minimum, reverses, and enters into a stable de
Sitter phase without a dramatic final. Also, we explore the aforesaid
cosmic scenarios by focusing on the leading contributions of some
physical quantities near the sudden future singularity and applying the
geometric Tipler and Kr\'olak criteria in order to  inspect the behavior
of timelike geodesic curves around such singularity.
\end{abstract}
\vskip 1cm \keywords{sudden future singularity, interaction,
variable vacuum energy, relaxed Chaplygin gas} \pacs{04.20.Dw,
98.80.-k, 98.80.Jk  }

\date{\today}
\maketitle
\section{Introduction}

Despite the overwhelming  observational evidence supporting the
current cosmic acceleration of the Universe coming from  supernovae
data, the cosmic microwave background anisotropies, and a
brand-new type of launched satellite for exploring the dark side of
gravity \cite{WMAP9,Planck2013}, quite little is known about the 
nature of dark energy (DE). More precisely, there is no a
fundamental theory at the microscopic level for explaning the origin of
DE, only some properties are known. For instance, astrophysical
observations suggest that DE is dominated by a strongly negative
pressure acting as a repulsive force \cite{WMAP9}. Another side of
this puzzle refers to what will be the ultimate fate of the Universe
\cite{ba}: could DE make  the Universe undergo an extremely
violent final event as a big rip singularity \cite{tipo1,tipo1b}, \cite{tipo1c}. 
This kind of singularity happens at a finite cosmic time where the
scale factor, the Hubble parameter, and its  cosmic time derivative
diverge \cite{tipo1b}.

Sudden future singularity is a one-of-a-kind future scenario that has
gained great interest recently \cite{tipo2,tipo2b} because it
offers an alternative and smooth  ``ultimate'' fate for the Universe,
opening the possibility for a noncatastrophic transition which can
lead to a new phase  in the Universe's evolution. The latter type of
doomsday happens at  a finite cosmic time where the scale factor and
the Hubble parameter remain finite but the  cosmic time derivative
diverges \cite{tipo2}. Indeed, the scale factor along with Hubble
parameter both remain bounded which implies the Christoffel symbols
are regular at this singularity. Hence, the geodesics are well
behaved and they can cross the singularity \cite{ruth}. Moreover, it
was found that a sudden future singularity does not experience
geodesic incompleteness \cite{barrow} and demonstrated that the
Tipler and Kr\'olak conditions do not hold \cite{tipler,krolak}; as
a result of that  finite objects (string or membrane) are
not crushed when crossing this singularity and, therefore, it can be
classified as a nonharmful transversable singularity. Also, it was
argued that the particles crossing the singularity will generate the
new geometry of the spacetime, providing in such a way a``soft
rebirth'' of the Universe after the singularity crossing \cite{ke}.
Relaxing one of the conditions that characterizes  a typical sudden
future singularity, another interesting
realization of sudden future singularity, dubbed  big brake
singularity \cite{gori}, was discovered that depends upon the Hubble parameter vanishing at a
finite cosmic time. The compatibility of this doomsday with
supernovae data was addressed in \cite{ke2}, and   an explicit
example of the crossing of this singularity was described in
\cite{ke},  where a tachyon field passes through the singularity,
continuing its evolution until  the Universe recollapses due to the
existence of final big crunch singularity.

Some appealing models  based on a  mixture of anti-Chaplygin gas
plus dustlike matter exhibiting a  soft (sudden/big brake) future
singularity were reported in the literature \cite{distribu}. In
fact,  the distributional version of anti-Chaplygin gas   was
presented  in Ref. \cite{distribu},  focusing in  the role played by  these
distributional quantities together with the junction conditions at
the singularity. Recently,  the issue of describing
a big-brake singularity in terms of a scalar field model was
analyzed in great detail in \cite{bbm}. It turned out that a big
brake future singularity can be obtained by a  modified Chaplygin
gas equation of state using a unified (effective) scheme.
Furthermore, this singularity was accommodated in terms of  an
exotic quintessence model and a full perturbation analysis was
carrried out near such event \cite{bbm}.

One of the aims of the present paper is to study some reasonable cosmological models with a finite-time future singularity when the Universe is filled with dark matter (DM) and variable vacuum energy (VVE) accommodated as dark energy. Contrary to the usual picture where the aforesaid  components are  decoupled, we propose that both fluids  exchange energy \cite{jefe2a}, \cite{jefe2b}, \cite{jefe2c}, \cite{jefe2d}, being the interaction a nonlinear combination of the total density and its first derivative. A physical motivation for selecting this kind of nonlinear interaction is that  the two interacting fluid model can be mapped into an effective unified model characterized by the relaxed version of the Chaplygin gas equation of state introduced in \cite{jefe1}. This equation of state produces an energy density that can be separated into three branches, one of them gives rise to a scale factor that interpolates between a matter dominated universe at early time and a de Sitter phase dominated by VVE at late times while the other two branches produce and energy density that vanishes o diverges at a finite scale factor value $a_s$. These are  attractive characteristics to investigate finite-time future singularities whenever $a_s=a(t_s)$ is finite for a finite cosmic time $t_s$. In doing so, we can provide a physical description of the behavior of the  interacting components near the singularity  but also we can compare with the behavior of effective quantities such as total energy and total pressure. In this way, we will be able to establish a physical correspondence between the critical behavior  of interaction and the effective equation of state.  Having mentioned the physical motivation of the nonlinear  interaction, we are going to use the so called ``source equation'' to determine the effective density, total pressure , and partial densities as well. Thus, We will reconstruct the explicit dependence of the interaction, total density, total pressure and partial densities in terms of the scale factor. Such procedure has the virtue of keeping the analysis simple but also it allows us to demonstrate the critical behavior of the total pressure, the interaction and the partial densities around the future singular event \cite{jefe3}. A remarkable fact is that the appearance of finite-time future singularity will be deeply connected with the critical behavior of the dark components near such event; that is,  the total energy density remains finite whereas the dark densities along with the total presssure will grow without limit.

The layout of the paper is as follows. In Sec. II, we introduce in detail
the model of an interacting dark sector composed of DM and VVE. In doing so,  we
solve the source equation, reconstruct all the geometrical, source
variables, and provide a classification of the different interaction
pieces in terms of the scale factor and the energy density in the
$(a,\ro)$ plane. This analysis is followed in  Sec. III by  solving the Friedmann equation and
exploring all the possible types of cosmic scenarios such as
big-bang singularity, bounce, big-crunch, sudden singularity, de
Sitter, and anti-de Sitter phases. Also, we study the behavior of
dark energy components: the energy density, the equation of state
and the stability of the de Sitter phases. In Sec. IV, we present
two piecewise nonlinear models obtained after matching two
interacting dark sector models at the finite-time future
singularity, which are characterized by a relaxed Chaplygin gas
equation of state, and describe the Universe's evolution identified
with both models. In Sec. V, we examine those singular events in
terms of comoving observers approaching to them by using the
criteria of Kr\'olak and Tippler. Section VI is devoted to present a
summary  of our main results.

\section{ Nonlinear interaction and finite-time future singularities in the FRW universe}

The purpose  of our paper is to examine the existence of finite-time
future singularities from the interaction  point of view. In doing so,  we 
investigate a two fluid model for a spatially flat FRW metric, where
the Universe is filled with DM and an unknown component that we
choose as VVE. The former one has a linear equations of state
$p_m=(\ga_m-1)\ro_m$, where the barotropic index $\ga_m$ is assumed
to be a constant such that  $\ga_m\simeq 1$,  while the latter one
has a vacuum equation of state $p_x=-\ro_x$ with $\ga_x=0$. The
energy density and the conservation equation of this system are
given by, \bn{ro} \ro=\ro_m+\ro_x, \ee \bn{co} \ro'=-\ga_m\ro_m, \ee
while the pressure  of the whole system is \bn{p} p=-\ro-\ro', \ee
and can be recast as $p=\ga_m\ro_m-\ro$. The prime stands for
derivatives with respect to the variable $\eta$ defined as $'\equiv
d/d\eta=d/3Hdt=d/d\ln{(a/a_0)^3}$, being $a_0$ some value of
reference for the scale factor $a$.  $H=\dot a/a$ is the expansion
rate and the dot  means $\dot{} \,\equiv d/dt $. Solving the linear algebraic
system of Eqs. (\ref{co}), one finds $\ro_m$ and $\ro_x$ as
functions of $\ro$ and its derivative $\ro'$: \be \n{rm}
\ro_m=-\frac{\ro'}{\ga_m}, \ee \bn{rx} \qquad
\ro_x=\ro+\frac{\ro'}{\ga_m}. \ee

We build a model of  interacting DM and DE by splitting the
conservation equation (\ref{co}) into two balance equations and
introducing the interaction term $Q$ with a factorized dependence as
$3HQ(\eta,\ro,\ro')$, \bn{qm} \ro'_m+\ga_m\ro_m=-Q, \ee \bn{qx}
\ro'_x=Q, \ee so there is an exchange of energy between DM and VVE
components. From Eqs. (\ref{rm}) and (\ref{qx}), we obtain the
``source equation'' \cite{jefe1} for  the total energy density \be
\n{se} \rho''+\ga_m\rho'= \ga_m Q. \ee Knowledge of the evolution of
$\ro$  requires to solve the  source equation (\ref{se}) for a given
$Q$. After having obtained $\rho$, we are in position to determine
the energy densities of DM and VVE components from Eqs. (\ref{rm})
and (\ref{rx}).

With the purpose of getting a solvable model and showing the
existence of finite-time future singularities in the interacting
dark sector, we present an interacting scenario generated  by a
nonlinear interaction in the form 
\bn{q} Q(\ro,\ro')=-n
\left[\rho'+\frac{\rho'^2}{\gamma_m(\rho-\rho_s)}\right], 
\ee 
where $\ro_s$ is a constant positive energy density and $n$ is the
coupling constant. Note that $Q$ diverges in the limit
$\ro\to\ro_s$. Let us mention which are the physical  motivations for selecting this ansatz over other choices. Cosmological scenarios where dark matter exchanges energy with a  modified holographic Ricci dark energy \cite{jefe2a}, \cite{jefe2b}, can be accommodated in terms of a nonlinear interaction, as that represented by Eq. (\ref{q}).  In dealing with that interacting framework,  one finds that the effective one-fluid obeys the equation of state of a relaxed Chaplygin gas. Therefore, the Universe is dominated by pressureless dark matter at early times and undergoes an accelerated expansion in the far future driven by a strong negative pressure \cite{jefe2a}. Another interesting case where the the nonlinear interaction takes place is related with the evolution of  a universe that has an interacting dark matter, a modified holographic Ricci dark energy plus a decoupled radiation term. The aforesaid model seems to be consistent with the Hubble data, the constraints coming from  the amount of dark energy during the recombination  along with the abundance of light elements obtained with  the big bang nucleosynthesis (BBN) data \cite{jefe2b}. 
 
Combining Eqs. (\ref{se}) and (\ref{q}), we obtain the
equation that governs the dynamics of the energy density, 
\bn{x''} x
x'' +\ga_m (n+1) x x'+n x'^2=0, 
\ee 
where a new dimensionless variable is defined as
 \be \n{x} x=\frac{\ro-\ro_s}{\bt\ro_s}, \ee where this positive variable is 
such that $x_s=x(\ro_s)=0$. Here the role played by the parameter
$\beta$ is the following: in the case $\bt=-1$ the energy density
and the variable $x$ range between $0\le\ro<\ro_s$ and $0<x\le 1$,
while for $\bt=1$ they define another physically admissible region
given by $\ro>\ro_s$ and $x>0$. Then, the energy density and its
derivative can be written as follows \bn{rr'} \ro=\ro_s(1+\bt x),
\qquad \ro'=\bt\ro_s x'. \ee Once the function $x=x(a)$ is known, we
calculate the interaction (\ref{q}), the energy density and its
derivative  (\ref{rr'}), the equation of state (\ref{p}) and the DM
and VVE densities (\ref{rm}) and (\ref{rx}) as functions of the
scale factor. To this end, we find the first integral of the source
equation (\ref{x''}) and its general solution which read, \bn{pi}
x'=-\ga_m[x+\al\,x^{-n}] , \ee \bn{xf}
x=\left\{\al\left[\left(\frac{a_s}{a}\right)^{3\ga_m(n+1)}-1\right]\right\}^{\frac{1}{n+1}},
\qquad n>-1, \ee where $\al$ is an integration constant. The
integration constant appearing in the general solution (\ref{xf}),
after integrating the first integral (\ref{pi}), was chosen so that
$x_s=x(a_s)=x(\ro_s)=0$ for a  finite value of the scale factor
$a_{s}\ne 0$, which is equivalent to demand the existence of finite
density at $a_s$, namely $\ro_s=\ro(x_s)=\ro(a_s)$. Such demand is
well justified provided we are seeking for  a realization of a
future singularity and it must have a nonvanishing Hubble function
at finite-time $t_{s}$ as we will see later on.  Besides, we must
combine $\al>0$ with $a<a_s$ or $\al<0$ with $a>a_s$ in order to
ensure that the square bracket in Eq. (\ref{xf}) is defined
positive. Finally,  if  Eqs. (\ref{pi}) and
(\ref{xf}) are combined, we  can obtain how  the first integral behaves with the scale factor, 
$x'=-\al\ga_m x^{-n}(a_s/a)^{3\ga_m(n+1)}$.

In the particular  case of a vanishing integration constant $\al$, the source equation
(\ref{x''}) of the interacting dark sector in presence of different
forms of dark matter components includes the $\Lambda$CDM model of
the General Relativity Theory. In fact, solving Eq. (\ref{pi}) and
inserting the solution into Eq. (\ref{rr'}), we obtain the energy
density \bn{rla}
\ro_{_{(\al=0)}}=\ro_s\left\{1\pm\left(\frac{a_0}{a}\right)^{3\ga_m}\right\},
\ee with $a_0$ an integration constant. From this equation we get
$\ro_{_{(\al=0)}}'=-\ga_m(\ro_{_{(\al=0)}}-\ro_s)$, so that
inserting it into Eqs. (\ref{p}), (\ref{rm}), (\ref{rx}) and
(\ref{q}), we have the equation of state
$p_{_{(\al=0)}}=-\ga_m\ro_s+(\ga_m-1)\ro_{_{(\al=0)}}$, which
reduces to $p_{_{(\al=0)}}=-\ro_s$ for cold dark matter(CDM) when
$\ga_m=1$, the dark energy densities $\ro_m=\ro_{_{(\al=0)}}-\ro_s$,
$\ro_x=\ro_s$ and a vanishing interaction term $Q_{_{(\al=0)}}=0$.
The upper sign in  Eq. (\ref{rla}) corresponds to the $\Lambda
$CDM model for $\ga_m=1$ and the lower one to a nonsingular
bouncing model.

Equations (\ref{x''})-(\ref{pi}) also admit a simple constant solution
$\ro_{_{dS}}=\ro_s[1+\bt(-\al)^{1/(n+1)}]$, which after having  used Eqs.
(\ref{rm})-(\ref{rx})  they lead to  $\ro_m=0$ and $\ro_x=\ro_{_{dS}}$.
Below, we will return to this solution that is related with the branches
of solutions which have an initial contracting or a final expanding
de Sitter phase. These stages  are in turn related  with  an asymptotically   vanishing
interaction. 

In order to study the most relevant outcome of our model, we collect
all the meaningful quantities which will be used later on. From Eqs.
(\ref{p})-(\ref{rx}), (\ref{q}) and (\ref{rr'})-(\ref{xf}), we find
that
$$
Q(a)=-n\al^2\bt\ro_s\ga_m\left(\frac{a_s}{a}\right)^{3\ga_m(n+1)}\,\,\,\,\,\,\,\,\,\,\,\,\,\,\,\,\,\,
$$
\bn{qf}
\,\,\,\,\,\,\,\,\,\,\,\,\,\,\,\,\,\,\times\left\{\al\left[\left(\frac{a_s}{a}\right)^{3\ga_m(n+1)}-1\right]\right\}^{-\frac{2n+1}{n+1}},
\ee
\bn{rf}
\ro=\ro_s\left\{1+\bt\left\{\al\left[\left(\frac{a_s}{a}\right)^{3\ga_m(n+1)}-1\right]\right\}^{\frac{1}{n+1}}\right\},
\ee
\bn{r'f}
\ro'=-\al\bt\ro_s\ga_m\left(\frac{a_s}{a}\right)^{3\ga_m(n+1)}\left\{\al\left[\left(\frac{a_s}{a}\right)^{3\ga_m(n+1)}-1\right]\right\}^{\frac{-n}{n+1}},
\ee
\bn{pf}
p=-\ro_{s}+(\ga_{m}-1)(\ro-\ro_s)+ \al\bt\ro_s\ga_{m}\left(\frac{\bt\ro_s}{\ro-\ro_s}\right)^n,
\ee
\bn{rmf}
\ro_m=\al\bt\ro_s\left(\frac{a_s}{a}\right)^{3\ga_m(n+1)}\left\{\al\left[\left(\frac{a_s}{a}\right)^{3\ga_m(n+1)}-1\right]\right\}^{\frac{-n}{n+1}},
\ee
\bn{rxf}
\ro_x=\ro_s\left\{1-\al\bt\left\{\al\left[\left(\frac{a_s}{a}\right)^{3\ga_m(n+1)}-1\right]\right\}^\frac{-n}{n+1}\right\}.
\ee

Now, we will focus our investigation  in the family of finite-time
future singularities such that  the scale factor and the Hubble
parameter remain finite at some finite cosmic time called $t_s$ but the
pressure diverges at this time. Meaning that the scale factor
$a_s=a(t_s)$, its first time derivative $\dot a_s=\dot a(t_s)$ are
finite but its second derivative 

\bn{as..} \ddot a_s=\ddot a(t\to
t_s)=\pm\,\infty, 
\ee 
diverging in the limit $t\to t_s$. As both the
scale factor and Hubble parameter remain bounded at $t=t_s$, the
Christoffel symbols are regular at this singularity and the
geodesics are well behaved and they can cross the singularity \cite{tipo2b}.
The existence of this kind of future singularity in the interacting
dark sector requires that the pressure (\ref{pf}), relaxed version
of the well-known Chaplygin gas, diverges in the limit
$\ro\to\ro_s$. Bearing this in mind and taking into account  Eq.
(\ref{pf}), we restrict the coupling constant to be positive $n>0$,
so that 
\bn{pl} p_s=p(\ro\to\ro_s)=\pm\infty, 
\ee 
as $\ro\to\ro_s$.Thus, $p\to-2\dot H\to -2\ddot a/a_s$ and the cosmic  acceleration
diverges $ \ddot a\to-a_sp_s/2\to\mp\,\infty$ when $\ro\to\ro_s$.
Also, the interaction $Q$, the $\eta$-derivative of the energy
density $\ro'$, the DM and VVE densities $\ro_m$ and $\ro_x$ diverge
in the limit $a\to a_s$ whereas their sign near $a_s$ is determined
by the sign of the product $\al\bt$. In this scenario, the pressure
$p=\ga_m\ro_m-\ro$ diverges as $a\to a_s$ provided the DM and VVE
densities $\ro_m\to\al\bt\,\infty$ and $\ro_x\to-\al\bt\,\infty$
blows-up at the limit $a\to a_s$  whereas the energy density remains
finite, namely $\ro=\ro_m+\ro_x\to\ro_s$. Our conclusion is that the
appearance of a finite-time future singularity is directly linked
with the behavior of the dark energy densities near the singularity
given that they diverge at $a_s$ together with the pressure as well.
Hence, the interaction between the dark components plays an important
role in the occurrence of the future singularity. This simple fact
shows a close relation between a divergent interaction term and
divergent dark energy components at $a_s$.

\begin{figure}[ht!]
\includegraphics[scale=0.582]{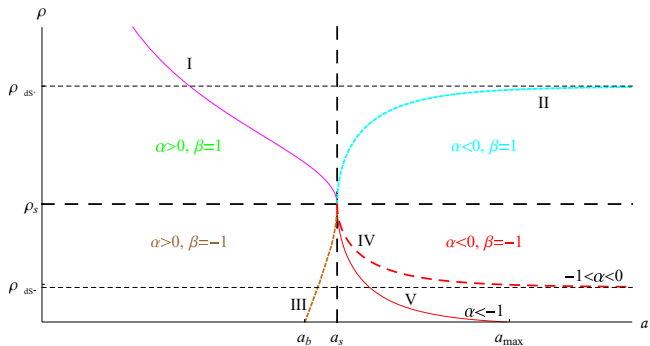}
\caption{Figure shows the energy density as a function of the scale factor for the four regions
characterized by the signs of $\al$ and $\bt$ parameters. Solutions
II and IV tend to constant values $\rho_{_{dS^+}}$ and
$\rho_{_{dS^-}}$, while solutions III and V reach
a bounce at $a_b$ and a maximum at $a_{max}$. In Sec. IV, we describe 
three different piecewise models
constructed by matching  solutions I with IV, I with V, and II
with III at $a=a_s$.
}
\label{Fig15}
\end{figure}

One way to address the full analysis  of this type of singular event
is to show more explicitly  how it can be  crossed  and by doing so
one  necessarily emerges in another kind of  universe. In order to
show how the matching of two nonlinear interacting dark sector can
be obtained, it is useful to start by classifying all the physical
distinctive dynamical regions of the model. Such a task is tackle by
considering a generic point $(a_s,\ro_s)$ in the plane $(a,\ro)$,
where $Q$, $\ro'$, $p$, $\ro_m$, $\ro_x$ and the effective
barotropic index $\ga=-\ro'/\ro$ diverge. The aforesaid analysis
points out that there are four regions in which the interaction term
and the others quantities change their specific forms (see for
example Fig. \ref{Fig15} for the energy density). These regions are
uniquely identified with the signs of the parameters $\al$ and $\bt$
(see Eqs. (\ref{x}) and (\ref{xf})). Clearly, the identification of
these four interaction pieces is characterized by the signs of $\al$
and $\bt$, so that we are going to define a new symbol
$Q^{{(\mbox{sign}(\al),~\mbox{sign}(\bt))}}$ for that purpose. The
four kinds of interactions can be classified in the following way: \bn{q++}
Q^{^{(+,+)}}(a)=Q(\al>0,\bt=1,a<a_s,\ro>\ro_s), \ee \bn{q--}
Q^{^{(-,-)}}(a)=Q(\al<0,\bt=-1,a>a_s,\ro<\ro_s), \ee \bn{q-+}
Q^{^{(-,+)}}(a)=Q(\al<0,\bt=1,a>a_s,\ro>\ro_s), \ee \bn{q+-}
Q^{^{(+,-})}(a)=Q(\al>0,\bt=-1,a<a_s,\ro<\ro_s). \ee In the
forthcoming sections we  are going to explore with enough detail two piecewise
models, one will be driven by interaction pieces
(\ref{q++})-(\ref{q--}) and the other will be associated with
(\ref{q-+}) and (\ref{q+-}), respectively.

\section{Solutions for the  Scale factor}

From the conclusions we have reached in  Sec.II, we are in a
position to assert that the Universe may evolve to a finite-time
future singularity.  We are going to take a closer look at  this
possibility by inspecting the physical mechanism behind  the
occurrence of such kind of event within the interacting dark sector
model. One way to achieve such a goal is examining not only the
source variables as energy density and pressure as functions of the
scale factor but also studying some of these quantities, for
example, the scale factor and the energy density as functions of the
cosmic time and its subsequent time derivatives provided both
variables are related through the Einstein's field equation. In this
direction, we will focus on the geometric part of the interacting
dark sector model and use the Friedmann equation, $3H^2=\ro$, to
describe the time evolution of the scale factor along with the
energy density. For that purpose, we rewrite the Friedmann equation in terms of the
dimensionless variable $x$ as
\be
\n{00} 
\dot x^2=3\ro_s(1+\bt x)\,x'\,^2.
\ee 
Equation (\ref{00}) suggests that we can find the approximate solution for the
scale factor along with the related quantities in  certain limiting
cases as a useful manner to find some connection between the
interacting dark sector and the existence of finite-time future singularities. 
We will examine how this procedure can be implemented in a consistent manner in the forthcoming
subsections.

\subsection{Scale factor solution near the sudden future singularity}

Near the singularity at $a=a_s$, where the interaction term (\ref{q}) diverges, $\ro\to\ro_s$ and $x\to 0$, we consider only the contribution of the term, $\al x^{-n}$, in the first integral (\ref{pi}) of the source equation (\ref{x''}) provided it is the leading term, so that the first integral reduces to the following one:
\bn{x'}
x'\approx -\al\ga_m x^{-n}.
\ee
Combining Eqs. (\ref{00}) and (\ref{x'}), we obtain the Friedmann equation near $a_{s}$,
\bn{x.f}
\dot x\approx -3\al\ga_mH_s\, x^{-n}\left[1+\frac{\bt}{2}x\right],
\ee
where the expansion rate $H_s=\sqrt{\ro_s/3}$ and whose approximate implicit solution is given by
\bn{s00}
\frac{x^{n+1}}{n+1}-\frac{\bt x^{n+2}}{2(n+2)}\approx-3\al\ga_m H_s(t-t_s).
\ee
Solving iteratively the last equation, we get the approximate time dependence of the scale factor, its first time derivative and the cosmic acceleration near the singularity at $t=t_s$; these are
\bn{a}
a(t)\simeq a_s\left[1+H_{s}\D t -\frac{ \bt\left[-3\al\ga_m(n+1)H_{s}\D t\right]^{\frac{n+2}{n+1}}}{6\al\ga_m(n+2)}+...\right],
\ee
\bn{a.}
\dot a(t)\simeq a_sH_s\left[1+\frac{\bt}{2}\left[-3\al\ga_m(n+1)H_{s}\D t)\right]^{\frac{1}{n+1}}+...\right],
\ee
\bn{a..}
\ddot a(t)\simeq-\frac{3\al\bt \ga_m a_sH^2_s}{2}\left[-3\al\ga_m(n+1)H_{s}\D t)\right]^{\frac{-n}{n+1}}+....,
\ee
where $\D t=t-t_{s}$. For $n>0$, we have that the scale factor $a(t_s)=a_s$ and its first time derivative $\dot a(t_s)=a_s H_s$ are both finite but its second time derivative $\ddot a\to-\,\al\bt\infty$ and the subsequent time derivatives diverge in the limit $t\to t_s$. When the parameters $\al$ and $\bt$ involved in the interacting piecewise model have the same signs a sudden future singularity will occur at the cosmic time $t_s$.

For the sake of completeness, we integrate the approximate first integral (\ref{x'}) and compose the solution with  the scale factor (\ref{a}),   in order to  find a power expansion of the energy density (\ref{rf}) as well as its time derivative near the singularity at $t_s$,
$$
\ro\approx\ro_s\left\{1+\bt\left[-3\al\ga_m(n+1)H_{s}\D t\right]^{\frac{1}{n+1}}\right. \,\,\,\,\,\,\,\,\,\,\,\,\,\,\,\,\,\,\,\,\,\,\,\,\,
$$
\bn{rex}
\,\,\,\,\,\,\,\,\,\,\,\,\,\,\,\,\,\,\left.\,\,+\frac{\bt^2\left[-3\al\ga_m(n+1)H_{s}\D
t\right]^{\frac{2}{n+1}}}{2(n+2)}+....\right\}, \ee \bn{r.ex}
\dot\ro\approx-3\al\bt\ro_s\ga_m H_s\left[-3\al\ga_m(n+1)H_{s}\D
t\right]^{\frac{-n}{n+1}}+...... \ee
Combining the cosmic
acceleration (\ref{a..}) with the first time derivative of the
energy density (\ref{r.ex}), we obtain a simple relation between
them 
\bn{r.ex'} \dot\ro\approx 6H_s\frac{\ddot a}{a_s}. 
\ee From
Eqs. (\ref{rex})-(\ref{r.ex'}),  we have explicitly a
finite energy density $\ro(t_s)=\ro_s$ at the singularity while its
time  derivative $\dot\ro\to-\,\al\bt H_s\,\infty$  in the limit $t\to t_s$ as long as 
the coupling constant remains positive ($n>0$).
In this case a sudden future singularity happens for an expanding
universe, $H_s>0$. In addition, from Eqs. (\ref{rm}) and (\ref{rx}),
we find that the dark energy densities and the barotropic index
$\ro_m\to\pm\infty$, $\ro_x\to\mp\infty$ and
$\ga=-\dot\ro/3H\ro\to\mp\infty$ diverge in the limit $t\to t_s$. At
the same time the interaction term $Q$ diverges at the cosmic time
$t_s$.

\subsection{Big Bang and Big Crunch singularities}

The interacting dark sector model has an initial singularity at $t=t_{_{BB}}$ when $\bt=1$, $\al>0$ and $n>-1$. In this case, the leading term in the energy density $(\ref{rf})$  is given by $\ro_{_{BB}}\approx\ro_s\al^{1/(n+1)}(a_s/a_{_{BB}})^{3\ga_m.}$, since  near the initial singularity $a_{_{BB}}\to 0$ and the energy density  blows up. Then, inserting $\ro_{_{BB}}$ into the Friedmann equation (\ref{00}), we find that the dominant term of the approximated solution is the power law scale factor,
\bn{sinf}
a_{_{BB}}(t)\simeq a_s\left [\frac{3}{2}\ga_m H_s\,\al^{1/2(n+1)} \,(t-t_{_{BB}})\right]^{\frac{2}{3\ga_m}}+.....,
\ee
with $t>t_{_{BB}}$ and $H_s>0$ for an expanding universe. We have that $\ro\to\ro_{_{BB}}\to\infty$ and $a\to 0$ when $t\to\tbb$  while the equation of state  (\ref{pf}) becomes linear  $p_{_{BB}}\approx(\ga_m-1)\ro_{_{BB}}$ and the pressure diverges in the same limit. Then, the big bang singularity occurs at $t=\tbb$, where the energy density $\ro$, the pressure $p$, $\dot\ro$, $\dot p$ and their higher derivatives diverge. Near this primordial singularity $\ro_m\to\infty$, $\ro_x\to\ro_s$, the interaction term (\ref{qf}) has a vanishing limit and dark components decouple in the limit  $t\to \tbb$. So, the appearance of the initial singularity is essentially caused by the growing without limit of the DM energy density. In contrast, at the sudden future singularity, the interaction $Q$
diverges and the dark components become strongly coupled. This is a
very significant result because the appearance of a finite-time
future singularity appears to be strongly linked with the divergence
of the interaction term at the limit $t\to t_s$. In the following,
we will reinforce such reciprocity.

Also there exists a possibility that a big crunch singularity occurs at  cosmic time $t_{_{BC}}=\tbb$ when the entire universe contracts from the past and collapses in the limit  $t\to t_{_{BC}}$, such that $t<t_{_{BC}}$ and $H_s<0$, under its own gravity until all known structures are concentrated at one point.

\subsection{Bouncing scenario}

For $\al>0$, $\bt=-1$ and $\ro\le\ro_s$ there exists a scale factor value $a_e$,
\bn{e}
a_e=a_s\left[1+\al^{-1}\right]^{\frac{-1}{3\ga_{m}(n+1)}},
\ee
such that the energy density (\ref{rf})  vanishes ($\ro_e=\ro(a_e)=0$) and the scale factor has an extremum. To explore the behavior of physical quantities near $a_e$,  it is useful  to define a small departure $\de$ from $a_e$ as follows
\bn{ep}
\de=\frac{a-a_e}{a_e}, \qquad  |\de|\ll 1.
\ee

Taking into account that $\ro_e=0$, we make a first order Taylor expansion of the energy density (\ref{rf}) around $a_e$,
\bn{}
\ro(\eta)\approx\ro'(\eta_e)(\eta-\eta_e)+....,
\ee
\bn{ro'e}
\ro'_e=\ro'(\eta_e)=(1+\al)\ro_s\ga_m,
\ee
where $\ro'(\eta_e)$ is positive definite, $\eta_e=3\ln {a_e}$ and $\eta-\eta_e=3\ln{a/a_e}=3\ln{(1+\de)}\approx 3\de$. The integration of  the approximate Friedmann equation, $\dot\de\approx \sqrt{\ro'(a_e)\de} $, leads to the scale factor in terms of the cosmic time, and it reads
\bn{ae}
a(t)\simeq a_e\left[1+\frac{\ro'_e}{4} \, (t-t_e)^2+....\right].
\ee
Then, $a_e$ represents a minimum of the scale factor where it bounces at $t_b=t_e$ and $a_b=a(t_b)=a_e$.

From Eqs. (\ref{qf})-(\ref{rxf}), we have that the interaction term (\ref{qf}) becomes constant at the bounce $Q(a_b)=-n\al\bt(1+\al)\ro_s\ga_m$, a similar result can be noticed for the DM and VVE densities $\ro_{mb}=\ro_{m}(a_b)=-(1+\al)\ro_s=-\ro_{xb}>0$ and the pressure $p_b=p(a_b)=-\ro'_b=-(1+\al)\ga_m\ro_s$ turns  negative. In short, all the source variables have a constant nonvanishing value at $a_b$. Such signature can be used as a physical property to distinguish a finite-time singularity from a bounce event. 

On the other hand, the energy density also vanishes if $\al<-1$, $\bt=-1$, and $\ro\le\ro_s$. The particular value $\al=-1$ is excluded from the analysis because the energy density vanishes in the limit $a\to\infty$. For the remaining values of $\al$, the extremum $a_e$ represents a maximum of the scale factor, meaning that the scale factor expands until it reaches the maximum value $a_{max}=a_e$ and then begins to contract (see Fig. (2)).

\begin{figure}[ht!]
\includegraphics[scale=0.87]{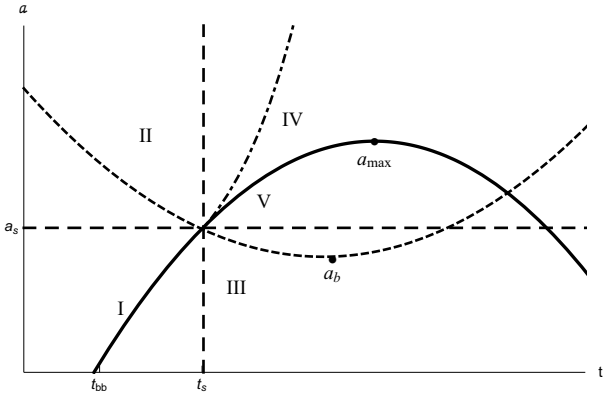}
\caption{Figure shows the qualitative behavior of the scale
factor as a function of the cosmic time obtained from the five 
energy densities of the Fig.~\ref{Fig15}. The
solid line represents the scale factor of the piecewise model
corresponding to energy densities I and V, the
dashed line to II and III,  and both the solid and the dot-dashed 
lines to I and IV.}
\label{Fig15A}
\end{figure}

\subsection{de Sitter phase}

In Sec. II, we have obtained the simple constant solution
\bn{rds}
\ro_{_{dS}}=\ro_s\left[1+\bt(-\al)^{1/(n+1)}\right], \quad  H_{_{dS}}={\sqrt{\frac{\ro_{_{dS}}}{3}}},
\ee
which includes a contracting de Sitter solutions for $\al<0$, $\bt=1$ and an expanding one for $-1<\al<0$, $\bt=-1$. For these solutions, we find that the dark energy densities are $\ro_{m_{dS}}=0$, and $\ro_{x_{dS}}=\ro_{_{dS}}$ with $p_{_{dS}}=-\ro_{_{dS}}$. In these cases,  the interaction term (\ref{q}) vanishes and the model becomes decoupled. We investigate the behavior of solutions $x=x_{_{dS}}(1+\ep)$ of Eq. (\ref{pi}) near to $x_{_{dS}}=\ro_s(1+\bt\ro_{_{dS}})$ with $\ep<<1$ by expanding the Eq. (\ref{pi}) to first order in $\ep$,
\bn{esds}
\dot\ep= -3(1+n)\ga_m H_{_{dS}}\,\ep.
\ee
This shows that in the regions defined by $\al<0$ and $\bt=\pm 1$, the contracting de Sitter phase (anti-de Sitter) $a\propto\exp{(-H_{_{dS}}t)}$ is unstable for $n>-1$ and the expanding one $a\propto\exp{(H_{_{dS}}t)}$ is stable. Then, the solution of the source equation (\ref{x''}) approach to the stable de Sitter solution (\ref{rds}) in the limit $t\to+\infty$.

\subsection{Comments on big brake and sudden future singularities}

In the last part of this section we will make some comments relative to the effective equation of state in the interacting dark sector and put in evidence of the main features of the big brake and the sudden future singularities when they occur as a product of nonlinear interacting processes. In particular, we will refer to two different cases which might seem to share  some similarities but they are  really different in nature as we will show later. On the one hand, we have the  the big brake singularity investigated in Ref. \cite{jefe3} when the dark sector includes two components, dark matter and VVE coupled with an interaction term having the form of Eq. (\ref{q}) with $\ro_s=0$. On the other hand, we have  the sudden future singularity examined in the present interacting model, where the nonlinear interaction term is given by  (\ref{q}) with $\ro_s\ne 0$. In the former case,  the effective equation of state has the form 
\bn{pb}
p_b=(\ga_{m}-1)\ro+ \al\ga_{m}\ro^{-n},
\ee
of a modified Chaplygin gas while in the latter case the effective equation of state is given by Eq. (\ref{pf}). Now, let us explore the essential differences between both equation of states which give rise to the  big brake and the sudden future singularities. To this end, we make use of the energy density expansion (\ref{rex}) near the future singularity at the cosmic time $t_s$ 
\bn{1}
\ro\approx\ro_s\left\{1+\bt\left[-3\al\ga_m(n+1)H_{s}\D t\right]^{\frac{1}{n+1}}+...\right\},
\ee
where without loss of generality, we have taken into account only the main contribution of the expansion (\ref{rex}) in power of $\D t$, and inserted it into the effective equation of state (\ref{pf}),
$$ 
p\approx -\ro_s+\bt\ro_s(\ga_m-1)[-3\al\ga_m(n+1)H_s\D t]^{\frac{1}{n+1}}
$$
\bn{2}
+\al\bt\ro_s\ga_m[-3\al\ga_m(n+1)H_s\D t]^{\frac{-n}{n+1}}+...
\ee
Surprisingly, the pressure $p\to 0$ in the limit $\ro_s\to 0$; hence, it does not reduce to equation of state (\ref{pb}) that characterizes the big brake singularity. However, in the case that $\ro_s\ne 0$, we have a sudden future singularity provided $1/(n+1)>0$ and $-n/(n+1)<0$, meaning that $n>0$. In fact, when $t\to t_s$ in Eqs. (\ref{a})-(\ref{a..}), we obtain that the scale factor $a\to a_s$ and its first time derivative $\dot a\to a_s H_s$ are finite while the acceleration $\ddot a\to\pm\infty$ diverges. Also,  taking into consideration  Eqs. (\ref{1}) and (\ref{2}) in the same limit, we have that the energy density $\ro\to\ro_s$ is finite but the pressure $p\to\pm\infty$ diverges. In conclusion, the big brake singularity cannot be obtained by taking the limit $\ro_s\to 0$ in the present model driven by the nonlinear interaction (\ref{q}). Therefore  we need to investigate the big brake and the sudden future singularities separately, being the former one identified with the nonlinear interaction (\ref{q}) with $\ro_s=0$ while the latter one must have  $\ro_s\ne 0$ due the physical definition of a sudden singularity.  The aforesaid  results are indicating  that the parameter $\ro_s$ plays a special role in describing a universe  with a sudden singularity, namely  $\ro_s$ introduces a discontinuity in the limit $\ro_s\to 0$ for the effective equation of state (\ref{q}) which is linked with the own nature of sudden singularity. In another words, the interacting model associated with the sudden singularity cannot lead continuously to the other interacting model related with the big brake singularity provided the own different physical nature of both singularities, so both scenarios deserve to  be explored  on their own merits. Indeed,  new physical insights can be gained by exploring this cosmological scenario and  because of that  many authors devoted several efforts to explore the emergence of the sudden singularity on general grounds or alternative frameworks \cite{tipo2, tipo2b}, \cite{barrow}. Here, we explored  other appealing aspects about the risen of sudden singularity along with the possibility of crossing this soft singularity within the interacting dark energy framework. Regarding the second topic, we will examine the joining of two universes connected by a sudden singularity in the next section  as a way to show the richness of the nonlinear interaction chosen in this work.       
 

\section{Piecewise nonlinear model in the dark sector}

So far we have presented an interacting dark-sector model in which DM interacts with VVE. The interaction has been divided in  four different pieces (\ref{q++})-(\ref{q--}) according to the four regions in the plane $(a,\ro)$ delimited by the sign of the parameters $\al$ and $\bt$, where  each one of these pieces has a physcial meaning  associated with  a particular region of that plane. In this section, we will extend our previous analysis on the cosmological model generated by those interaction pieces, in particular, we will match two different kinds of interaction pieces in order to show a possible extension of the Universe through the finite-time future singularity at $t=t_s$. In doing so,  we will examine the interaction pieces near the future singularity, explore the behavior of the energy density, its $\eta$ derivative, the DM, and VVE densities along with the equation of state of the content of the Universe. In a way, we will offer a connection  between two universes (one of these prior to the singularity event) and another universe (posterior to that event) emerging from the future singularity, taking into account that in both cases the content of the Universe includes interacting DM and VVE. Also, the interacting model will produce several types of singularity events and different kinds of  cosmological scenarios that we will investigate in detail.

\subsection{ Cosmological model driven by \\ $\qmm$ and $\qee$}

As we mentioned earlier, the present interacting cosmological model
has two parts; one is driven by $\qmm(a)$ which is  defined by
$\al>0$, $\bt=1$ and turns out to be restricted to the region  $a<a_s$
and $\ro>\ro_s$  in the plane $(a,\ro)$.  However, there is another
part  driven by $\qee(a)$,  which is  specified by $\al<0$, $\bt=-1$,
so it is associated with the region $a>a_s$ and $\ro<\ro_s$. The
explicit dependence on the scale factor for  $\qmm$ and  $\qee$
together with other useful quantities of the model are listed in
Eqs. (\ref{qf})-(\ref{q+-}).

For $n>0$,  the interaction piece $\qmm<0$ and the Universe starts
from a big bang singularity at $t=\tbb$, where $\qmm\to 0$,
$\ro\to\infty$, $\ro'\to-\infty$, the pressure $p\to\infty$, the DM
energy density $\ro_m\to\infty$, and the VVE density $\ro_x\to\ro_s$
in the limit $a\to 0$.  Later, the the model describes a matter-dominated 
universe with a power-law scale factor. Interestingly
enough, the DM density decreases until it  reaches its minimum value
at $\ro_{m,{ {min}}}=\ro_s(1+n^{-1})(\al n)^{1/(n+1)}$, where
$\ro_m'=0$ and $\ro$ has an inflection point. Then, the DM density
begins to increase and finally $\ro_m\to +\infty$ when $a\to a_s$.
However, the VVE density at the initial singularity is
$\ro_x=\ro_s$, later decreases with the scale factor provided
$\ro_x'=\qmm<0$, so then vanishes and,  finally, $\ro_x\to-\infty$ as
$a\to a_s$ while $\ro=\ro_m+\ro_x\to\ro_s$ remains finite.

Near the cosmic time $t_s$, the approximate scale factor (\ref{a}) takes the
following form
$$
\amm\simeq a_s\left[1+H_{s}\D t-\frac{\left[-3\al\ga_m(n+1)H_{s}\D t\right]^{\frac{n+2}{n+1}}}{6\al\ga_m(n+2)}...\right],
$$
\bn{a++}
\al>0,\,\,\,\bt>0,\,\,\,a<a_s,\,\,\,\ro>\ro_s,\,\,\,H_s>0,\,\,\,t<t_s.
\ee The sudden future singularity occurs at the finite-time $t_s$
where the scale factor $a_s=a(t_s)$ and its first time derivative
$\dot a_s=\dot a(t_s)=a_sH_s$ are both finite  but its
second time derivative $\ddot a\to-\infty$ for  $t\to t_s$. The
Hubble expansion rate $H_s=H(t_s) $ and the energy density
(\ref{rex}), $\ro_s=\ro(t_s)$, has a finite value at the
singularity; however, the pressure $p=\ga_m\ro_m-\ro\to+\infty$ diverges
in the limit $t\to t_s$.

In the other part of the piecewise model driven by the positive interaction piece $\qee>0$, the scale factor (\ref{a}) near $a_s$ can be recast as
$$
\aee\simeq a_s\left[1+H_{s}\D t+\frac{\left[-3\al\ga_m(n+1)H_{s}\D t\right]^{\frac{n+2}{n+1}}}{6\al\ga_m(n+2)}...\right],
$$
\bn{a--}
\al<0,\,\,\,\bt<0,\,\,\,a>a_s,\,\,\,\ro<\ro_s,\,\,\,H_s>0,\,\,\,t>t_s.
\ee The DM and VVE densities (\ref{rmf}) and (\ref{rxf}) have the
limits $\ro_m\to+\infty$ and $\ro_x\to-\infty$, respectively, when
$a\to a_s$. However, the energy density $\ro=\ro_s$ remains finite
although the pressure $p=\ga_m\ro_m-\ro\to+\infty$ diverges. In
addition, we have that the scale factor and its first time
derivative $a_{s}$ and $\dot a_s=a_sH_s$ are finite whereas the
cosmic acceleration $\ddot a\to-\infty$ in the limit $t\to t_s$.

Let us make a comment about the final piecewise model based on the
process  of matching two interacting universe. The scale factor and
expansion rate ({\it geometrical variable}) along with the DM and
VVE densities, the energy density, its time derivative, and pressure
of the effective fluid ({\it source variables}) have the same limits
in the limit $t\to t_s$ independently whether the future singularity
is reached by the first  part of the model which we have identified
with $\qmm$ or by the second part of the model identified with
$\qee$. So that, the evolution across the singularity at $t_s$ is
completely regular  which in turn  means that comoving observers can transit from an
expanding universe driven by $\qmm$, passing through the sudden
future singularity,  and emerging in an expanding universe driven by
$\qee$ where the singularity now is located  in  its  distant past.

Far away from the singularity, the solutions of the model driven by
$\qee$ evolve in two different forms according to the values
assigned to the integration constant $\al$. For instance, when $
\al<-1$ there is a set of scale factor solutions for which the
energy density vanishes $\ro_0=\ro(a_{max})=0$ at a specific value
of the scale factor \bn{amax}
a_{max}=a_s\left[1+\al^{-1}\right]^{\frac{-1}{3\ga_{m}(n+1)}}, \ee
which is the maximum value reached by the scale factor (\ref{ae})
belonging to that set of solutions, and its expression near
$a_{max}=a(t_{max})$ is given by \bn{ae'} a(t)\simeq
a_{max}\left[1+\frac{(1+\al)\ro_s\ga_m}{4} \,
(t-t_{max})^2+...\right]. \ee After that, the Universe reverses its
expansion ($H_s<0$) and begins to collapse, possibly into a big
crunch. In this case, it seems that the catastrophic final of the
universe identified with the present piecewise model would be unavoidable
anyway.

For the other set of solutions, for which the integration constant
ranges between $-1<\al<0$, the energy density always decreases and
varies between $\ro_{_{dS}}<\ro<\ro_s$. In this case, this expanding
branch of solutions emerge from a singularity in the past and the
interaction piece $\qee$ accelerates the expansion, driving the
universe to a final de Sitter stage described by the Eq. (\ref{rds})
with $\bt=-1$. From Eqs. (\ref{rmf}) and (\ref{rxf}), we have that
$\ro_m\to 0$, the VVE density increases asymptotically to achieve
the final energy density $\ro_{x}\to\ro\to\ro_{_{dS}}$ and the
asymptotic equation of state turns in
$p_{_{dS}}\approx-\ro_{_{dS}}$, being the fuel that controls the
final dynamic of the Universe. Consequently,  the interaction piece
$\qee\to 0$ at late times and the dark sector becomes decoupled in
the remote future.

Summarizing, we have constructed a piecewise model driven by the
interaction pieces $\qmm$ and $\qee$. This model describes a
universe that begins in the far past from a big bang singularity in
the region of the plane ($a,\ro$) identified with $\qmm$. Then it
has a matter-dominated era and exhibits an accelerated expansion
until it reaches the sudden future singularity at $t_s$, where the DM
and VVE densities $\ro_m\to+\infty$, $\ro_x\to-\infty$ and the
pressure $p=\ga_m\ro_m-\ro$ diverge in the limit $t\to t_s$. This
clearly shows that the DM and VVE densities produce the sudden
future singularity. In the other region of the piecewise model, the
interaction piece $\qee$generates two branches of solutions; one
expanding branch emerges from a distant past singular event, with the
scale factor reaching a maximum from which the Universe reverses and
collapses into itself, possibly in a big crunch with a catastrophic
finale. However, the other expanding branch ends in a stable de
Sitter phase, avoiding a dramatic final fate.

\subsection{ Cosmological model driven by \\ $\qem$ and $\qme$}

We will construct a different piecewise model by gluing two interacting models. One part of the model is
based on the options $\al<0$ and  $\bt=1$ so that $a>a_s$ and  $\ro>\ro_s$, indicating that the exchange of energy is driven by
the interaction piece $\qem(a)$. On the other hand,  the choice $\al>0$ and $\bt=-1$ corresponds to
$a<a_s$ and $\ro<\ro_s$ which is related with the interaction piece $\qme(a)$. The main quantities involved in this piecewise model, as
functions of the scale factor, are obtained from Eqs.
(\ref{qf})-(\ref{q+-}).

The Universe begins with a contracting de Sitter (inflationary)
phase in the far distant past, with a decoupled dark sector,
$\qem\approx 0$, $\ro_m\approx 0$,
$\ro\approx\ro_x\approx\ro_{_{dS}}$ and equation of state
$p_{_{dS}}\approx-\ro_{_{dS}}$, until it nearly reaches the cosmic
time  $t_s$ where the approximate scale factor (\ref{a}) is given by
$$
\aem\simeq a_s\left[1+H_{s}\D t -\frac{\left[-3\al\ga_m(n+1)H_{s}\D t\right]^{\frac{n+2}{n+1}}}{6\al\ga_m(n+2)}...\right],
$$
\bn{a-+}
\al<0,\,\,\,\bt>0,\,\,\,a>a_s,\,\,\,\ro>\ro_s,\,\,\,H_s<0,\,\,\,t<t_s.
\ee From this equation, we have that both $a\to a_s$ and $\dot
a\to\dot a_s=a_s H_s<0$ remain finite at $t=t_s$, while the cosmic
acceleration grows without ($\ddot a\to+\infty$) at the limit
$t\to t_s$. This points  out that the piecewise model also has a finite-time future
singularity at $t_s$. Regarding the dark energy components, we find
that $\ro_m\to-\infty$, $\ro_x\to+\infty$ while  $\ro\to\ro_s$,
$\ro'\to+\infty$ and $p\to-\infty$ when $t\to t_s$. In addition, the
energy density ranges between  $\ro_{_{dS}}\le\ro\le\ro_s$, whereas
the scale factor varies from $\infty$ to $a_s$.

When the Universe emerges from the finite-time future singularity at
$t_s$, the other interaction piece $\qme$ takes the control of the
dynamics, leading to a contracting phase. Consequently, the scale
factor (\ref{a}) is written as
$$
\ame\simeq a_s\left[1+H_{s}\D t+\frac{\left[-3\al\ga_m(n+1)H_{s}\D t\right]^{\frac{n+2}{n+1}}}{6\al\ga_m(n+2)}...\right],
$$
\bn{a+-}
\al>0,\,\,\,\bt<0,\,\,\,a<a_s,\,\,\,\ro<\ro_s,\,\,\,H_s<0,\,\,\,t>t_s.
\ee
The interaction piece $\qme$ and the remaining relevant quantities $\ro$, $\ro'$, $p$, $\ro_m$, $\ro_x$ as functions of the scale factor are given by Eqs. (\ref{qf})-(\ref{rxf}) together with (\ref{q+-}). In this piece of model, $\qme\to+\infty$, $\ro(a_s)=\ro_s$, $\ro'\to+\infty$ and the pressure $p\to-\infty$, causing $\ddot a\to+\infty$ as $t\to t_s$. Taking into account  that $\ro_m<0$ and  $\ro_x>0$ in the aforesaid scenario, we find that $\ro_m\to-\infty$ and $\ro_x\to+\infty$ while the sum of the partial densities remains bounded $\ro=\ro_m+\ro_x=\ro_s$ in the limit $t\to t_s$. These outcomes are in agreement with our previous results obtained for {\it geometrical and source variables} which characterize a finite-time future singularity.

The Universe continues as before until bounces at the scale factor value $a_b$, given by
\bn{ab}
a_b=a_{s}\left[1+\al^{-1}\right]^{\frac{-1}{3\ga_{m}(n+1)}},
\ee
for which $\ro(a_b)=0$, $a_b\le a\le a_s$ [see Eq. (\ref{e})] and the approximated scale factor (\ref{ae}) near the bounce at $t=t_b$ takes the next form
\bn{ab'}
a(t)\simeq a_b\left[1+\frac{(1+\al)\ro_s\ga_m}{4} \, (t-t_b)^2+...\right].
\ee

The Universe reverses its contraction at $t=t_b$, begins to expand ($H_s>0$) and possible will enter into a stable de Sitter phase driven by the fuel provided by the VVE density which dominates over the DM one.

Summing up, the piecewise model driven by the interaction pieces $\qem$ and $\qme$ gives rise to a universes which begins in the distant past in a contracting de Sitter phase, then one can extend the evolution of the Universe through the finite-time future singularity, emerges from that singularity and  reaches a bounce until reverses it. Then the Universe begins to expand and the aforesaid dynamical mechanism  produces a stable de Sitter phase described by the asymptotic equation of state of the dark sector (\ref{pf}),  $p_{_{dS}}\approx-\ro_{_{dS}}$, indicating that  the VVE component becomes the dominant contribution.

\section {Tipler and Kr\'olak method }

In the previous sections we extracted quite general conditions for
the existence of a finite-time future singularity within the
framework of an interacting dark sector based on a given nonlinear
interaction and a piecewise nonlinear interaction counterpart. We
elaborated some new appealing cosmic scenarios by matching in a
smooth way two different interacting sectors with the idea of
showing how the finite-time future singularity can be continued,
offering a new scheme where the aforesaid singularity is viewed as a
two-door way to another phase of the Universe.  We have also mentioned
why this ansatz is physically relevant provided that the effective
model is related with an relaxed version of the Chaplygin gas
equation of state. In this section, we will try to focus on a
geometrical aspect of the previous findings. Thus,  we would like to
complete our analysis of the future singularity by taking into
account a geometric point of view  based on a method developed by
Tipler \cite{tipler} and Kr\'olak \cite{krolak}. To summarize these
criteria briefly, we recall that  a spacetime is Tipler strong
\cite{tipler}  if as the proper  time   $t\rightarrow t_{s}$, the
integral \bn{ti} {\cal
{T}}(t)=\int^{t}_{0}dt'\int^{t'}_{0}{|{\cal{R}}_{\al\beta}u^{\al}u^{\beta}|}dt''
\rightarrow \infty. \ee In same manner,  a spacetime is Kr\'olak
strong \cite{krolak}  if as the proper  time  $t\rightarrow t_{s}$,
the integral \bn{ki} {\cal
{K}}(t)=\int^{t}_{0}{|{\cal{R}}_{\al\beta}u^{\al}u^{\beta}|}dt
\rightarrow \infty, \ee where the components of the Ricci tensor are
understood to be written in a frame which is transported parallel
along the geodesic curves.  A  singularity or event can be strong by
Kr\'olak criteria but weak according to Tipler's criteria, however,
the reverse situation always holds. Because weak singularities can
be extended beyond them, the method developed by Tipler and Kr\'olak
are  useful tools for determining the fate of the Universe in terms
of the fate of geodesic curves near potential strong singular point
\cite{comment}. Let us consider  timelike geodesic curves, $x^{i}=c$
with $i$ spatial index and $c$ a constant \cite{geo}, associated
with comoving observer; i.e.,  we take into account a comoving
worldline congruence with velocity
$u^{\alpha}=(\partial_{t})^{\alpha}=(1, 0,0,0)$ so that the proper
time coincides with the coordinate time. The components of the Ricci
tensor measured by  the obsever along this congruence  lead to
${\cal{R}}_{~\al\beta}u^{\al}u^{\beta}=-3\ddot{a}(t)/a(t)$
\cite{inva}. Let us take into account the first piecewise model.
Using  (\ref{a++}), calculating its  second  derivative, and
replacing in Eq. (\ref{ki}), we can demonstrate that  the Kr\'olak
invariant yields ${\cal {K}}(t) \propto [\kappa (t -t_{s}) ]^{m}$
with $m>0$ provided the exponent satisfies the relation
$(n+2)/(n+1)>1$, so the Krolak measure does not diverge as
$t\rightarrow t_s$; such fact indicates that the sudden event is
K-weak.  On the other hand, the  Tipler measure involves a second
integration, giving  a ${\cal {T}}(t) \propto [\kappa (t -t_{s})
]^{m+1}$ regular quantity as $t\rightarrow t_s$ because $n>-1$; this
shows that the aforesaid  event is  T-weak also and, therefore, a
transversable singularity. The same kind of procedure can be done
for the second part of the piecewise model with (\ref{a--}). The
regularity of  these invariants on the both sides of the singularity
ensures the continuity of the limits at the sudden future
singularity. This fact is closely related with the regularity of the
scale factor at both sides of the sudden future singularity. We
obtained similar results for the other piecewise  model associated
with  (\ref{a-+}) and (\ref{a+-}). Besides, the weakness of sudden
future singularity was recently proved  \cite{barrow}.

As a short comment, let us mention that at the bounce the  Tipler
and Kr\'olak measures behave in a regular way because
$\ddot{a}(\eta)/a(\eta) \propto [1-p\eta^2]$ with $\eta= t -t_{b}$
while for the big bang or big crunch singularities, both invariants
diverge.

\section{Summary}

We have investigated an interacting dark sector model composed of DM
and VVE with linear equations of state for a spatially flat FRW
spacetime and obtained the source equation associated with this  system. We have solved the source
equation and obtained the general solution for the dark sector energy
density, when the DM and VVE are coupled with an extension 
of the nonlinear interaction introduced in Ref. \cite{jefe1}. The
general solution includes several family of solutions, so with the purpose of
constructing a finite-time future
singularity at $t=t_s$  within  this interacting framework, we have selected the family of solutions which
produces a divergence of the nonlinear interaction for
$\ro\to\ro_s=\ro(a(t_s))$, where $t_s$ is an specific value of the
cosmic time such that the scale factor $a_s=a(t_s)$ and the energy
density $\ro_s=\ro(a_s)$ are both finite at the limit $t\to t_s$. 
For this family of solutions, we have shown that the cosmic
acceleration $\ddot a/a=\ro/3+\ro'/2\to\pm\,\infty$ because $\ro'$
diverges as $a\to a_s$. Additionally, the pressure of the whole dark
sector $p=-\ro-\ro'$, that is related with a relaxed version of the
Chaplygin gas equation of state, diverges in the limit $\ro\to\ro_s$
for $t\to t_s$. In this interacting dark sector model, the pressure
$p=\ga_m\ro_m-\ro=(\ga_m-1)\ro-\ga_m\ro_x$, diverges as a
consequence that both DM and VVE densities also diverge in the limit
$t\to t_s$. In the same manner, the occurrence of a finite-time future
singularity is strongly associated with the divergent behavior of
$\ro_m$ and $\ro_x$. It was shown the importance of considering a non
linear coupling between them in order to show how the behavior of dark energy densities 
 produces that singularity. After simple considerations,
we have separated the nonlinear interaction term in four pieces
according to the four regions in the plane ($a,\ro$), defined by the
scale factor, the energy density and uniquely identified with the
signs of the parameters $\al$ and $\bt$. In each one of these
regions the nonlinear interaction changes its form and has a proper
significance. In this  context, we have obtained the solution of the
Friedmann equation for the scaled factor which describes several
types of singularities and different kinds of scenarios which we
have investigated in enough detail. In particular, by using the
approximate scale factor we have studied its behavior near the
finite-time future singularity, big bang and big crunch as well as
power law, bounce, and de Sitter scenarios.

We have constructed a piecewise cosmological model driven by the
interaction pieces $\qmm$ and $\qee$,  gluing the corresponding scale
factors at $t_s$ to extend the Universe through the sudden future
singularity located at $t_s$. We have explored the dynamics of the model  by examining
the behavior of the energy density, its derivative, the DM and VVE
densities along with the equation of state of  the Universe. We have described an expanding
matter-dominated universe that starts from a big bang singularity
driven for $\qmm$ and then continues to a sudden future singularity.
We have encountered that  the scale factor $a_s=a(t_s)$, its first
time derivative $\dot a_s=\dot a(t_s)$ and the energy density
$\ro_s=\ro(a(t_s))$ are finite when they are evaluated at $t_s$, but the
acceleration $\ddot a$ diverges for $t\to t_s$. Regarding  the dark 
densities, we have found that  $\ro_m\to+\infty$, $\ro_x\to-\infty$ while the total pressure
$p=\ga_m\ro_m-\ro$ diverge in the limit $t\to t_s$. This showed us  the
strong relation that exists between sudden future singularity at
$t_s$ and divergent behavior of DM and VVE densities for $t\to
t_s$. After that, the Universe emerged from a singularity in the past
driven by $\qee$ with two possible branches of solutions. The branch
$\al<-1$, reaches a maximum at
$a_{max}=a(t_{max})>a_s$, then reverses its expansion, collapses and
possibly end in a big crunch singularity. Nevertheless,  the case with 
$-1<\al<0$ represents an expanding
scenario and the Universe evolves asymptotically towards an
accelerated de sitter phase. These solutions are asymptotically
stable and avoid a dramatic final fate.

We have generated a second piecewise model with the interaction pieces $\qem$ and $\qme$, and gluing the respective scale factor solutions at $t_s$. The Universe begins in the distant past in a contracting de Sitter phase, with a nearly vanishing interaction piece $\qem\approx 0$ and a decoupled dark sector. The DM energy density $\ro_m\approx 0$,  the energy density $\ro\approx\ro_x\approx\ro_{_{dS}}$ and the VVE energy has an approximate de Sitter equation of state $p_{_{dS}}\approx-\ro_{_{dS}}$. The Universe contracts and then continues to a finite-time future singularity at $t=t_s$, where the scale factor $a_s=a(t_s)$, its first time derivative $\dot a_s=a_s H_s<0$ and the energy density remain finite at $t=t_s$, while the cosmic acceleration grows without limit ($\ddot a\to+\infty$) as  $t\to t_s$, so these results indicate a finite-time future singularity at $t_s$. Then, the Universe emerges driven by the interaction piece $\qme$, bounces at the scale value $a_b=a(t_b)$, and after that it reverses, begins to expand, and subsequently the VVE density begins to dominate over the DM one. Then, the expanding universe has a stable de Sitter phase, described by the asymptotic dark sector equation of state,  $p\to p_x\to p_{_{dS}}\approx-\ro_{_{dS}}$. Notice that the approximate scale factor can be extended at the finite-time future singularity with the one obtained from the interaction piece $\qme$ at $t=t_s$.

Using the approximate scale factor in the vicinity of this singular event along with the geometric method developed by Tipler and Kr\'olak for the case of timelike geodesic curves  (comoving observer), we have obtained  a regular behavior of Tipler and Krolak measures  around the sudden future singularity; that is, both measures are well behaved at both sides of the singularity and matched continuously at $t_s$, which in turn ensures the regularity of these measures for the piecewise model in the whole spacetime. In short, timelike geodesic curves can be extended beyond such singular event, turning into a traversable singularity.

In the near future,  we will  implement an analysis of the classical
stability associated with the sudden future singularity for the
piecewise models which will be mainly focused on the scalar modes of
the perturbed metric. Such analysis will complement the already
known  stability studies of sudden future singularity \cite{bl}.

\acknowledgments

This work was supported by Grants MEC-CONICYT No. 80140060 (M. C. and L.P.C), FONDECYT-CONICYT N$^0$ 1140238 (M.C.) and Direcci\'on de Investigaci\'on de la Universidad del B\'{\i}o--B\'{\i}o through grants No. 140807 4/R and No. GI 150407/VC (M..C.). L. P. C. thanks to UBA under Project No. 20020100100147, CONICET under Project PIP 11420110100317, and acknowledges the hospitality of the Physics Department of Universidad del B\'{\i}o--B\'{\i}o. M.G.R. is supported by Conselho Nacional de Desenvolvimento Cient\'{\i}fico e Tecnol\'ogico (CNPq), Brazil. 



\begin{thebibliography}{99}
\bibitem{comment}
 The analysis performed with  a causal geodesic  which meets  a $T/K$-strong singularity  gave the same kind of finding because it involves the  integral  of some components of the Riemann tensor which are  parallel transported, namely  ${\cal{R}}^{\al}_{~\beta\gamma\delta}u^{\beta}u^{\delta}$, for which the nonvanishing components turned out to be $ {\cal{R}}^{\al}_{~t\gamma t}=-\ddot{a}/a$ also \cite{ruth}.
\bibitem{geo}
In a FRW spacetime, curves of the form $x^{i}(t)=c$, with $t$ the proper time fulfill the geodesic equation, $\ddot{x}^{\alpha}+\Gamma^{\alpha}_{~\beta\nu}u^{\beta}u^{\nu}=0$, provided
the only component of 4-velocity is $u^{t}=1$, $\dot{u^{t}}$=0,  and  $\Gamma^{t}_{~tt}=0$.
\bibitem{inva}
For a FRW metric,   the components of the Ricci tensor are  ${\cal R}_{00} =-3(H^2+\dot{H})$ and ${\cal R}_{ij} =a^3(H^2+\dot{H})\gamma_{ij}$ with $\gamma_{ij}$ the spatial metric of the three-dimensional submanifold, whereas  the Ricci scalar is  ${\cal R}=6(H^2+ \frac{\ddot{a}}{a})$.
\end{thebibliography}
\end{document}